\documentclass[twocolumn, pra]{revtex4}
\usepackage{amssymb}

\begin{document}
\input epsf

\title{The physics of no-bit-commitment : Generalized quantum non-locality
versus oblivious transfer}

\author{Tony Short$^{a}$, Nicolas Gisin$^b$, Sandu Popescu$^{a,c}$}
\affiliation{ $^a$HH Wills Physics Laboratory, University of
Bristol, Tyndall Avenue, Bristol, BS8 1TL, UK\\  $^b$ Group of
Applied Physics, University of Geneva, 20 rue de
l'Ecole-de-m\'{e}decine, CH-1211 Geneva 4,
Switzerland.\\
$^c$Hewlett-Packard Laboratories, Stoke Gifford, Bristol BS12 6QZ,
UK}

\begin{abstract}
We show here that the recent work of Wolf and Wullschleger
(quant-ph/0502030) on oblivious transfer apparently opens the
possibility that non-local correlations which are  stronger than
those in quantum mechanics could be used for bit-commitment. This
is surprising, because it is the very existence of non-local
correlations which in quantum mechanics prevents bit-commitment.
We resolve this apparent paradox by stressing the difference
between non-local correlations and oblivious transfer, based on
the time-ordering of their inputs and outputs, which prevents
bit-commitment.

\end{abstract}

\maketitle

In 1984, Bennett and Brassard \cite{bb84} proposed a quantum
physics scheme (BB84) by which two parties can establish a secret
key, allowing them to communicate with unconditional security
against eavesdroppers. This result remains one of the cornerstones
of quantum cryptography, and gave rise to the hope that many other
cryptographic primitives (which classically rely on unprovable
assumptions), could be made unconditionally secure within a
quantum framework. Perhaps the best known of these is \emph{bit
commitment}.

A bit-commitment scheme allows one party (Alice) to commit to a
decision in such a way as a second party (Bob) will believe her
when she later reveals it, but cannot find out her decision until
that point (eg. Alice's decision is sealed in a safe which is
given to Bob, while Alice keeps the key. Then in the revealing
stage, Alice sends the key to Bob).

Using the same encoding scheme as for key-distribution, Bennett
and Brassard constructed a quantum coin-tossing protocol
\cite{bb84} that could directly be used for bit-commitment. Their
protocol is secure when Alice is limited to using only separable
states. However, as noted by the authors, Alice can cheat
convincingly in their protocol by using entangled states.

This result was later expanded by Mayers \cite{quantum_bc1}, and
by Lo and Chow \cite{quantum_bc2}, to show that Alice can cheat in
\emph{any} quantum bit-commitment scheme which is secure against
Bob. It is therefore impossible to implement an unconditionally
secure bit-commitment scheme within quantum physics.

Entangled quantum states are crucial in proving this
\emph{no-bit-commitment} result, allowing correlations to exist
between Alice's and Bob's systems which cannot be simulated by any
local hidden-variable model \cite{bell}. These `non-local'
correlations do not allow for super-luminal signalling, but are
nevertheless extremely powerful \cite{qi}.

Interestingly, Popescu and Rohrlich have shown that quantum states
do not provide the strongest possible non-local correlations
consistent with relativity \cite{popescu}. For two parties, who
both have a single binary input (their measurement setting) and a
binary output (their measurement result), the strongest possible
correlations are instead given by systems known as PR-boxes
\cite{popescu, barrett, singlet-sim}.

PR-boxes are a valuable conceptual tool in understanding
non-locality, as they allow us to separate the concept of
non-local non-signalling correlations from the details of a
particular physical model (eg. Complex Hilbert space).

PR-boxes (and their analogues with more inputs and outputs) are
also very general. All bipartite no-signalling boxes with binary
inputs and outputs can be constructed from a PR-box and a mixture
of local operations \cite{barrett}. Furthermore, a single PR-box
(with shared randomness) can be used to simulate the results of
any bipartite measurement on a maximally entangled quantum singlet
state \cite{singlet-sim}.

Recently, Wolf and Wullschleger \cite{wolf} have shown that a
PR-box can also be used to simulate a cryptographic primitive
known as \emph{one-out-of-two oblivious transfer} \cite{OT1, OT2},
in which Bob can secretly learn either (but not both) of two bits
submitted by Alice. We will refer to the device which implements
this scheme as an OT-box.

Interestingly, it is known that OT-boxes \emph{can} be used to
implement secure bit commitment. If OT-boxes can be built from
PR-boxes, then this implies that bit commitment can be achieved
using PR-boxes. This would be a surprising result, as PR-box
correlations are very similar to those attainable in quantum
theory, where, as emphasized previously, no bit-commitment is
possible.

In order to investigate this argument in more detail, this letter
is organized as follows. First, we introduce the requirements for
a bit commitment scheme and give an explicit example of such a
scheme using OT-boxes. Next, we recall Wolf and Wullschleger's
connection between the PR- and the OT-boxes. Finally, we show that
the analogous scheme involving PR-boxes \emph{does not} in fact
allow secure bit-commitment and elaborate on its significance.

Formally, a Bit commitment scheme consists of two protocols
(COMMIT and REVEAL) which satisfy the following requirements:
\begin{enumerate}
\item \textbf{Correctness}: If Alice and Bob are both honest, then
during the COMMIT protocol, Alice selects a value for her
committed bit $\alpha \in \{0,1\}$, and this value is learnt by
Bob during the REVEAL protocol.

\item \textbf{Privacy}: If Alice is honest, then Bob can learn nothing about $\alpha$ until the
REVEAL protocol is enacted.

\item \textbf{Binding}: If Bob is honest, then after the COMMIT protocol has finished,
there is (at most) only one specific value of $\alpha$ (e.g.
$\alpha=0$) which Bob will accept during the REVEAL protocol. He
will never accept the other possible value of $\alpha$. This
prevents Alice from `changing her mind' about which value of
$\alpha$ she committed.
\end{enumerate}

A bit-commitment scheme satisfying these three conditions would be
\emph{perfectly secure}. However, for cryptographic purposes it is
interesting to consider the slightly weaker case of \emph{secure}
bit commitment, in which  a bit-commitment scheme can be made
arbitrarily close to perfectly secure.

In particular, we consider a weakening of the binding requirement
to the following:

\begin{enumerate}
\setcounter{enumi}{2}
\item \textbf{Secure binding}: The scheme must permit arbitrarily large
values of a security parameter $N_{\epsilon} \in \mathbb{N}$, such
that if Bob is honest, then after the COMMIT process has finished
there is at most one value of $\alpha$ (e.g. $\alpha=0$) which Bob
will accept with probability greater than $2^{-N_{\epsilon}}$ in
the REVEAL protocol. This is Alice's committed value of $\alpha$.
\end{enumerate}

Note that given a secure bit-commitment protocol with secure
binding parameter $N_{\epsilon}=1$, it is possible to obtain a
protocol with an arbitrarily large security parameter
$N_{\epsilon}'$ by running $2N_{\epsilon}'-1$ copies of the
$N_{\epsilon}=1$ protocol in parallel (with Bob only accepting
Alice's commitment if she is not caught cheating in any of the
parallel runs). As the schemes we consider have perfect
\emph{privacy}, this will not help Bob gain any further
information about Alice's committed bit.

\begin{figure}
\begin{tabular}{lclc}
&\epsfxsize=1.5truein \epsffile{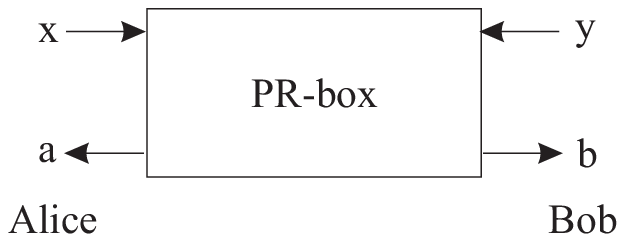} &  &
\epsfxsize=1.5truein\epsffile{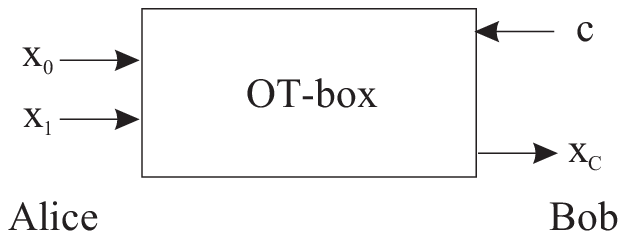}
\end{tabular}
 \caption{A schematic representation of the PR-box (left), and the OT-box (right)
 showing the input and output parameters for Alice and Bob.}  \label{pr_ot_fig}
\end{figure}

Now, it is possible to construct a secure bit-commitment scheme
using \emph{oblivious transfer}. Following Wolf and Wullschleger
\cite{wolf}, we consider OT-boxes which implement one-out-of-two
oblivious transfer, in which Alice inputs two bits, $x_0$ and
$x_1$, into the system and Bob inputs a single choice bit $c$.
Finally, the system outputs $x_c=x_0 \oplus c(x_0 \oplus x_1)$ to
Bob, where $\oplus$ denotes addition modulo 2. This allows Bob to
learn one of Alice's two input bits, without Alice knowing which
bit he has obtained (see figure \ref{pr_ot_fig}).

A well-known way to implement bit-commitment from one-out-of-two
oblivious transfer is to use this to simulate ordinary oblivious
transfer \cite{crepeau} and then use the latter for bit commitment
\cite{kilian}. Here we give an explicit protocol which uses
one-out-of-two oblivious transfer (our OT-box) to directly
implement bit-commitment.

Consider first a scheme in which Alice and Bob share a single
OT-box, which is as follows:
\begin{description}
\item[COMMIT]:
\begin{enumerate}
\item Alice selects a random bit $r$, and her committed bit $\alpha$.
\item Alice inputs $x_0 = r$ and  $x_1 = r \oplus \alpha$ into the OT-box,
where $\oplus$ denotes addition modulo 2.
\item Alice sends a message to Bob telling him it is his turn.
\item Bob selects a random bit $s$ and inputs $c = s$ into the
OT-box
\item Bob records the output bit $x_c$. If the OT-box fails to produce an
output (as Alice has not yet made her inputs), Bob knows that
Alice is cheating and will not accept her revealed bit.
\end{enumerate}

\item[REVEAL]:
\begin{enumerate}
\item Alice sends $\alpha$ and $r$ to Bob.
\item Bob checks to see if  $x_c = r \oplus \alpha s$. If this relation is true,
Bob accepts $\alpha$ as the revealed bit, otherwise he knows that
Alice has cheated and rejects Alice's revelation.
\end{enumerate}

\end{description}

COMMIT and REVEAL constitute a secure bit-commitment scheme with
security parameter $N_{\epsilon}=1$, as can easily be checked.
Using $2N'_{\epsilon}-1$ OT-boxes in parallel, it is therefore
possible to obtain arbitrarily secure bit commitment.

We now consider whether one can generate an analogous
bit-commitment scheme to that given above using PR-boxes (figure
\ref{pr_ot_fig}). Recall that a PR-box can be thought of as an
abstraction and generalization of a Bell-type experiment
\cite{bell}, in which Alice and Bob share an entangled system on
which they can each perform one of two different dichotomic
measurements. If we denote Alice's and Bob's binary inputs (their
measurement settings) by $x$ and $y$ respectively, and their
binary outputs (their measurement results) by $a$ and $b$
respectively, the extremal non-local correlations given by a
PR-box are of the form
\begin{equation} \label{pr_eqn}
a \oplus b = x y,
\end{equation}
where all outcomes consistent with (\ref{pr_eqn}) are equally
likely, and each party obtains their output bit immediately after
entering their input. Although such correlations are stronger than
anything attainable in quantum theory, Alice and Bob cannot use
the PR-box to signal to one another.

In order to simulate a PR-box using an OT-box and
\emph{vice-versa}, Wolf and Wullschleger propose the following
protocols \cite{wolf}:

\begin{description}
\item[PR-box from OT-box]:
\begin{enumerate}
\item Alice chooses a random bit $a$.
\item Alice inputs $x_0 = a$ and $x_1= x \oplus a$ into the OT-box.
\item Bob inputs $c=y$ into the OT-box and obtains output $b=x_c$.
Note that $b = a \oplus x c = a \oplus x y$ as required.
\end{enumerate}

\item[OT-box from PR-box]:
\begin{enumerate}
\item Alice inputs $x=(x_0\oplus x_1)$ into the PR-box and obtains output $a$.
\item Alice sends $m=x_0 \oplus a$ to Bob
\item Bob inputs $y=c$ into the PR-box and obtains output $b$.
\item Bob computes the output $x_c=m \oplus b$. Note that $x_c=m \oplus b
=x_0 \oplus (a \oplus b) = x_0 \oplus c (x_0 \oplus x_1)$ as
required.
\end{enumerate}
\end{description}

Directly applying the latter simulation procedure to the COMMIT
protocol for the OT-box, we find the analogous procedure COMMIT$'$
for the PR-box. The REVEAL protocol is the same in both cases:

\begin{description}
\item[COMMIT$'$]:
\begin{enumerate}
\item Alice selects a random bit $r$,
and her committed bit $\alpha$.
\item Alice inputs $x= \alpha$ into the PR-box, and
records her output bit $a$
\item Alice sends Bob the message $m = r \oplus a$
\item Bob selects a random bit $s$ and
inputs $y = s$ into his PR-box.
\item Bob records his output bit-string $b$, and computes $x_c = b \oplus m$.
\end{enumerate}
\end{description}

Apparently the correctness, privacy and secure binding of this
protocol COMMIT$'$ should be just as good as for the protocol
COMMIT (i.e. secure bit commitment with $N_{\epsilon}=1$), and
should therefore yield arbitrarily secure bit-commitment if enough
PR-boxes are used. However, this strongly suggests that a similar
protocol based on quantum entanglement instead of the PR-boxes
should work equally well, something proven impossible!

In fact, the PR-box protocol does not provide secure bit
commitment, as we will now show. As is the case when proving the
impossibility of quantum bit commitment, it is the secure binding
that fails. Indeed, unlike the OT case, Bob has no way of telling
if Alice has applied her inputs during the COMMIT$'$ protocol, as
the PR-box will give him an output as soon as he applies his
input, regardless of Alice's actions. Interestingly, this allows
Alice to successfully cheat by adopting the following strategy:
During the COMMIT$'$ protocol, Alice does not select her committed
bit $\alpha$ or apply any input to her PR-box, but instead sends a
random bit $m$ to Bob. Then, during the REVEAL protocol, Alice
chooses which $\alpha$ she wishes to reveal, enters $x= \alpha$
into her PR-box, and transmits $r= a \oplus m$ (as well as her
selected $\alpha$) to Bob. When he checks whether he should accept
$\alpha$, Bob will find that
\begin{equation}
x_c = b \oplus m =r \oplus a \oplus b= r \oplus \alpha s,
\end{equation}
and will therefore accept Alice's revealed $\alpha$ with
certainty, despite the fact that it was selected \emph{after} the
COMMIT$'$ protocol had finished. This violates the secure binding
condition, which requires that only one value of $\alpha$ will be
accepted with probability greater than $1/2$ during the REVEAL
protocol (for $N_{\epsilon}=1$). Even with multiple PR-boxes in
parallel, Alice can use the same trick independently on each to
cheat with certainty. COMMIT$'$ and REVEAL do not therefore form a
secure bit-commitment protocol.

Let us stress that all PR-boxes must yield an output for Bob even
if Alice has not yet entered her inputs, as this is essential to
prevent signalling from Alice to Bob. This is in full analogy with
the case of quantum entanglement. For instance, if Alice and Bob
share a singlet, then Bob can measure his half independently of
Alice, and vice-versa. There is thus no chance of Bob detecting
Alice cheating during the COMMIT$'$ protocol.

Consequently, despite the fact that it is possible to simulate the
outputs of an OT-box using a PR-box and 1 bit of communication,
this result shows that the simulation is not universally
composable. In particular, the simulation cannot be used to
implement a secure bit-commitment protocol in the same way as the
original OT-box.

The crucial difference between the OT-box and the PR-box lies in
the time ordering of the inputs. Bob can only obtain an output
from the OT-box after Alice has entered her inputs, and can
therefore check to see whether Alice has entered her inputs during
the COMMIT protocol. In the case of the PR-box, the output is
independent of the time-ordering of the two measurements, and no
such check exists. \footnote{Furthermore, in the case of the
PR-box, Alice has an output which is correlated with Bob's input
and output, which she can use to help her cheat.}

In conclusion, in \cite{wolf} Wolf and Wullschleger showed that
there is a deep connection between a fundamental primitive of
\emph{non-local correlations}, the PR-box, and a fundamental
cryptographic primitive, \emph{oblivious transfer}. In particular
they show that a PR-box can be used to simulate the correlations
produced by an OT-box and vice-versa.

In this letter, we have shown that such simulation is not the
whole story. In addition to recreating the correct probability
distributions, it is important to incorporate any restrictions on
the timings of the box inputs. Such restrictions play a crucial
role in any attempt to implement secure bit-commitment schemes
using the two primitives.

We have shown that the protocols COMMIT and REVEAL form a secure
bit-commitment scheme using OT-boxes. However, the analogous
protocols COMMIT$'$ and REVEAL, in which the OT-boxes are
simulated using PR-boxes, do not constitute a secure
bit-commitment scheme as they are not binding.

The main issue is that, due to the non-signalling character of the
PR box, Alice can postpone her measurements, and actually perform
them during the REVEAL protocol, rather than the COMMIT$'$
protocol. We emphasize that it is this liberty (which is in common
with quantum mechanics) that is the main element in the
no-bit-commitment property of nature, rather than any of the
particular characteristics of quantum entanglement.

The above results were obtained for a particular bit-commitment
scheme based on the OT-box protocol. However, we conjecture that
no-bit-commitment is a general feature of nature in the presence
of non-signalling non-local correlations, since any such
correlations allow Alice the liberty of postponing her
measurements.  Note that in order for this conjecture to be true,
we need to allow for the existence of all possible non-local
correlation-boxes (not only PR boxes) \cite{barrett}, and the
corresponding dynamics of such boxes (which generalize quantum
evolutions) \cite{dynamics}. Any limitations could allow
bit-commitment. This is similar to the quantum mechanical case: If
Alice can only use limited entanglement, it is easy to construct a
reliable protocol based on states for which cheating would require
more entanglement than Alice can produce.

To conclude, no signalling - no obligation to perform the
measurement until the end- no commitment!

\textbf{Note added}: After the completion of this work, we became
aware of a work by Buhrman et al. \cite{buhrman} that addresses
the same problem from a different perspective.

\begin{acknowledgments}
The authors acknowledge support from the U.K. Engineering and
Physical Sciences Research Council (IRC ``Quantum Information
Processing'') and from the E.U. under European Commission project
RESQ (contract IST-2001-37559).
\end{acknowledgments}

\end{document}